\documentclass[review]{elsarticle}
\journal{Journal of \LaTeX\ Templates}

\newcommand {\sqrts}{$\sqrt{s}$ }

\begin{document}

\begin{frontmatter}

\title{Measurement of forward photon production cross-section in proton--proton collisions at $\sqrt{s}$~=~13~TeV with the LHCf detector}

\author[INFN-Fi,Univ-Fi]{O.~Adriani}
\author[INFN-Fi,Univ-Fi]{E.~Berti}
\author[INFN-Fi]{L.~Bonechi}
\author[INFN-Fi,Univ-Fi]{M.~Bongi}
\author[INFN-Fi,Univ-Fi]{R.~D'Alessandro}
\author[Polytech]{M.~Haguenauer}
\author[Nagoya-ISEE,Nagoya-KMI]{Y.~Itow}
\author[Waseda]{T.~Iwata}
\author[Waseda]{K.~Kasahara}
\author[Nagoya-ISEE]{Y.~Makino}
\author[Nagoya-ISEE]{K.~Masuda}
\author[Nagoya-ISEE]{E.~Matsubayashi}
\author[Nagoya-Sci]{H.~Menjo}
\cortext[mycorrespondingauthor]{Corresponding author}
\ead{menjo@isee.nagoya-u.ac.jp}
\author[Nagoya-ISEE]{Y.~Muraki}
\author[INFN-Fi]{P.~Papini}
\author[INFN-Fi,CNR-Fi]{S.~Ricciarini}
\author[Nagoya-ISEE,Nagoya-KMI]{T.~Sako}
\author[Tokushima]{N.~Sakurai}
\author[Nagoya-ISEE]{M.~Shinoda}
\author[Waseda]{T.~Suzuki}
\author[Kanagawa]{T.~Tamura}
\author[INFN-Fi,Univ-Fi]{A.~Tiberio}
\ead{alessio.tiberio@fi.infn.it}
\author[Waseda]{S.~Torii}
\author[INFN-Ca,Univ-Ca]{A.~Tricomi}
\author[Berkeley]{W.C.~Turner}
\author[Nagoya-ISEE]{M.~Ueno}
\author[Nagoya-ISEE]{Q.D.~Zhou}

\address[INFN-Fi]{INFN Section of Florence, Florence, Italy}
\address[Univ-Fi]{University of Florence, Florence, Italy}
\address[Polytech]{Ecole-Polytechnique, Palaiseau, France}
\address[Nagoya-ISEE]{Institute for Space-Earth Environmental Research, Nagoya~University, Nagoya, Japan }
\address[Nagoya-KMI]{Kobayashi-Maskawa Institute for the Origin of Particles and the Universe, Nagoya~University, Nagoya, Japan}
\address[Waseda]{RISE, Waseda University, Shinjuku, Tokyo, Japan }
\address[Nagoya-Sci]{Graduate School of Science, Nagoya~University, Nagoya, Japan }
\address[CNR-Fi]{IFAC-CNR, Florence, Italy}
\address[Tokushima]{Tokushima University, Tokushima, Japan }
\address[Kanagawa]{Kanagawa University, Kanagawa, Japan }
\address[INFN-Ca]{INFN Section of Catania, Italy}
\address[Univ-Ca]{University of Catania, Catania, Italy}
\address[Berkeley]{LBNL, Berkeley, California, USA}

\begin{abstract}
In this paper, we report the production cross-section of forward photons in the pseudorapidity regions of $\eta\,>\,10.94$ and $8.99\,>\,\eta\,>\,8.81$, measured by the LHCf experiment with proton--proton collisions at \sqrts = 13~TeV. The results from the analysis of 0.191 $\mathrm{nb^{-1}}$ of data obtained in June 2015 are compared to the predictions of several hadronic interaction models that are used in air-shower simulations for ultra-high-energy cosmic rays. 
Although none of the models agree perfectly with the data, EPOS-LHC shows the best agreement with the experimental data among the models. 
\end{abstract}


\end{frontmatter}

%\linenumbers

%%%%%%%%%%%%%%%%%%%%%%%%%%%%%%%%%%%%%%%%%%%%%%%%%%%%%%%%%%%%%%%%%%%
\section{Introduction}
Hadronic interaction models play an important role in ultra-high energy cosmic-ray (UHECR) observations.
They are used in Monte Carlo~(MC) simulations of air-shower developments induced by UHECRs, which are one of the key tools used for reconstructing information about primary cosmic rays from observables measured by ground-based detectors.
Currently, the Pierre Auger Observatory~\cite{ref:PAO} and Telescope Array~\cite{ref:TA} are taking data for UHECRs. Although the experiments have published the results of the measured observables which are sensitive to the chemical composition of UHECRs, they have not yet reached any clear conclusions, because of the uncertainty related to the choice of hadronic interaction model~\cite{ref:PAO_Xmax,ref:PAO_muDepth,ref:TA_Xmax}.  
Since it began operation in 2009, the Large Hadron Collider (LHC), the world's largest hadron collider, has provided unique opportunities for testing hadronic interaction models with collision energies exceeding $10^{15}$ eV in a fixed target frame (\cite{ref:d'Enterria} for review of early results).
The major models used in air-shower simulations for UHECRs were re-tuned and updated by taking into account several experimental results obtained from proton--proton collisions with center-of-momentum collision energies of 0.9 and 7~TeV. These models, QGSJET~II-04~\cite{ref:QGSJET2}, EPOS-LHC~\cite{ref:EPOS}, and SIBYLL~2.3~\cite{ref:SIBYLL}, are called the post-LHC models. 
However, even with these post-LHC models, inconsistencies between the observed data and MC simulations were reported~\cite{ref:PAO_model}. 

The LHC forward (LHCf) experiment~\cite{ref:TDR}, one of the LHC experiments designed to test hadronic interaction models, was running during the early phase of the LHC operation with proton--proton collisions at \sqrts = 13~TeV in 2015. 
In this paper, we report the results of photon analyses performed on the taken data.  
The production cross-section of photons, of which 90\% are decay products of $\pi^{0}$ mesons produced in collisions, is analyzed in two pseudorapidity ranges.
The results of the photon analyses for the lower-energy collisions of \sqrts = 0.9 and 7~TeV have been published in Ref.~\cite{ref:900GeV-photon,ref:7TeV-photon}.  
Because of a collision energy nearly a factor of two higher than 7 TeV, the collision energy in the fixed target frame, $0.9 \times 10^{17}$ eV, was about a factor of four higher and the coverage of the transverse momentum $p_{\mathrm{T}}$ of the measurement was a factor of two wider than that at \sqrts = 7~TeV.  

The LHCf possesses two sampling and imaging calorimeter detectors which are installed on both sides of the LHC interaction point IP1~\cite{ref:Detector}. Each of the two detectors, Arm1 and Arm2, has two calorimeter towers with acceptances of 20~mm $\times$ 20~mm and 40~mm $\times$ 40~mm (Arm1), and 25~mm $\times$ 25~mm and 32~mm $\times$ 32~mm (Arm2). 
This double-calorimeter configuration allows the photon pairs to be detected from the decay of $\pi^{0}$ and $\eta$ mesons with threshold energies of 600\,GeV and 2.2\,TeV, respectively.
Each calorimeter consists of 16 scintillator layers interleaved with 44 radiation lengths of tungsten plates. 
The detectors are located 140\,m from IP1 and, in nominal operation, the smaller towers cover a pseudorapidity ($\eta$) range above 10, including the zero-degree collision angle. 
The larger towers are located above the smaller towers oriented $0^\circ$ (Arm1) and $45^\circ$ (Arm2) in the clockwise direction from the vertical.    
They cover the slightly off-center region where 8.5 $<\,\eta\, <$ 9.5.  
Before operation in 2015, the detectors were upgraded to improve their radiation hardness by replacing the plastic scintillators with $\mathrm{Gd_{2}SiO_{5}}$ (GSO) scintillators~\cite{ref:GSO}.
Four pairs of X-Y scintillating-fibre hodoscopes used in the Arm1 imaging sensor were also replaced with X-Y GSO bar-bundle hodoscopes~\cite{ref:GSObarTest}. 
In addition, four X-Y pairs of silicon detectors inserted in the Arm2 detector were upgraded to optimise the linearity. The performance of the upgraded detectors was studied in two beam tests at CERN-SPS before and after operation at the LHC. We confirmed that the energy and position resolutions for electromagnetic showers were better than the requirements of $<\,5\%$ and $<\,200\,\mathrm{\mu m}$, respectively~\cite{ref:BeamTest}.   

In this paper, we present the forward photon production cross-section in two regions of photon pseudorapidity ($\eta > 10.94$ and $8.81 < \eta < 8.99$) measured by the LHCf detectors.
All photons with energies above 200 GeV produced directly in collisions or from subsequent decays of directly produced short-lifetime particles (i.e. particles with $c \cdot \tau < 1$ cm, where $c$ is the speed of light and $\tau$ is the mean lifetime of the particle) are considered.

%%%%%%%%%%%%%%%%%%%%%%%%%%%%%%%%%%%%%%%%%%%%%%%%%%%%%%%%%%%%%%%%%%%
\section{Data}

The experimental data used in this analysis were obtained by a dedicated LHCf run from 22:32 to 1:30 (CEST) on June 12--13, 2015, during proton--proton collisions at \sqrts\,=\,13\,TeV. This operation period corresponds to the first three hours of LHC Fill 3855, which was one of the low-luminosity LHC runs operated with fewer bunches and a higher $\beta^{*}$ of 19 m than the LHC's nominal condition.   
In the Fill, 29 bunches collided at IP1 with a half crossing angle of 145~$\mu\mathrm{rad}$. In addition, six and two non-colliding bunches at IP1 circulated in the clockwise and counter-clockwise beams, respectively.  The total luminosity of the colliding bunches during data acquisition was measured by the ATLAS experiment at $L\,=\,(3-5)\,\times\,10^{28}\,\mathrm{cm^{-2}s^{-1}}$~\cite{ref:ATLAS-luminosity}. The number of collisions per bunch crossing, $\mu$, was in the range of 0.007--0.012.
% and the expected pile-up rate of inelastic collisions was  between 0.3 \% and 0.6 \%. 
Considering an acceptance of the detectors of about 15\% for inelastic collisions, the pile-up of events on a detector was negligible in this analysis. 

The recorded total integral luminosity was 0.191 $\mathrm{nb^{-1}}$ after correction of the data-acquisition live time. 
%Assuming the inelastic cross-section of $\sigma_{\mathrm{inela}}$\,=\,78.5\,\rm{mb}, it corresponds to $1.50\times10^{7}$ inelastic collisions. 
The numbers of recorded shower events in Arm1 and Arm2 were 1.79 and 2.10 M, respectively. 
The trigger efficiency was 100\% for photons with energies greater than 200 GeV.

%%%%%%%%%%%%%%%%%%%%%%%%%%%%%%%%%%%%%%%%%%%%%%%%%%%%%%%%%%%%%%%%%%%
\section{MC Simulation}
\label{sec:mc}
A full MC simulation was performed to obtain some parameters and correction factors used in this analysis and to validate the analysis method. The simulation consisted of the following three parts: 1) event generation of p--p inelastic collisions at IP1; 2) particle transportation from IP1 to the front of the detector; and 3) detector response. 
All three parts were implemented with MC simulation packages Cosmos~7.633~\cite{ref:Cosmos} and EPICS~9.15~\cite{ref:EPICS}. In the first part of the simulation, either QGSJET~II-04 or EPOS-LHC was used as an event generator and the DPMJET~3.04~\cite{ref:DPMJET3} model was used as a hadronic interaction model in the detector simulation of the third part. We generated $10^{8}$ inelastic collisions with the QGJSET~II-04 model. The dataset was used as a template sample for particle identification (PID) correction and a training sample for the unfolding method described in Sec.~\ref{sec:Corrections}. Another full MC simulation dataset of $5 \times 10^{7}$ inelastic collisions was generated with EPOS-LHC and used to validate the analysis method and estimate systematic uncertainties.

In addition, we generated $10^{8}$ events of inelastic p--p collisions with each hadronic interaction model, EPOS-LHC, QGSJET~II-04, DPMJET~3.06, SIBYLL~2.3, and PYTHIA~8.212~\cite{ref:PYTHIA}, using either the PYTHIA dedicated generator or CRMC~1.6.0, an interface tool of event generators~\cite{ref:CRMC}.
The decay of short-lifetime particles with $c \cdot \tau$ less than 1 cm was treated in these generators.
These event sets were used only in Sec.~\ref{sec:Results} to compare the photon production cross-section of the data and model predictions.
The total inelastic cross-section predicted by each model was used to express the results as the differential cross-section ($d\sigma/dE$).
The cross-sections used for each model are listed in Table~\ref{tab:cross}. 

\begin{table}[!htbp]
  \begin{center}
    \begin{tabular}{|c||c|c|c|c|c|}
      \hline
      \textbf{Model} & EPOS & QGSJET & DPMJET & SIBYLL & PYTHIA\\
      \hline
      \boldmath$\sigma_{inel}$ \textbf{[mb]} & 78.98 & 80.17 & 80.14 & 79.86 & 78.42\\
      \hline
    \end{tabular}
    \caption{Total inelastic cross-section ($\sigma_{inel}$) for a p--p collision at 13 TeV predicted by each hadronic interaction model (version number is omitted for simplicity).}
    \label{tab:cross}
  \end{center}
\end{table}

%%%%%%%%%%%%%%%%%%%%%%%%%%%%%%%%%%%%%%%%%%%%%%%%%%%%%%%%%%%%%%%%%%%

\section{Analysis}
\subsection{Event Reconstruction}

In this analysis, we used an event reconstruction algorithm resembling that employed in Ref.~\cite{ref:7TeV-photon,ref:Operaiton2010}.
The detector upgrades warranted a revaluation of the calibration parameters by beam tests~\cite{ref:GSObarTest,ref:BeamTest}. Then, the criteria in this analysis were re-optimised by MC simulation studies. 
We selected the events that met the criteria of PID for photons and the rejection of multi-hit events in which two or more particles hit a calorimeter tower.

The reconstructed energy of each event was rescaled by the factor obtained from a study of $\pi^{0}$ events, in which photon pairs were detected by the two calorimeter towers of each detector. The invariant mass of a photon pair was calculated using both the measured photon energies and hit positions, assuming the decay vertex coincides with IP1. The distribution of the reconstructed mass had a peak corresponding to the $\pi^{0}$ mass. 
We compared the peak masses from the data and the MC simulations and obtained energy rescale factors of $+3.5$\% and $+1.6$\% for Arm1 and Arm2, respectively. 
The factors were consistent with the systematic uncertainty of energy-scale calibrations discussed in Sec.~\ref{sec:sys_scale}. 

In this analysis, we defined two analysis regions, A and B. Region A is the area of a half-disk shape with $R < 5$~mm and $\Delta\phi = 180^\circ$, where $R$ is the distance from the beam center and $\Delta\phi$ is the azimuthal interval on each detector plane. 
The beam center was defined as the projection of the beam direction at IP1 on the detector surface. 
Region B is the sector-shape area for which $35 \mathrm{mm} < R < 42 \mathrm{mm}$ and $\Delta\phi = 20^\circ$. Regions A and B correspond to the pseudorapidity regions $\eta\,>\,10.94$ and $8.81\,<\,\eta\,<\,8.99$, respectively. Only the events for which the reconstructed hit positions are within these two regions were used in the final results.
The azimuthal acceptance was then corrected in the final results.
A position resolution of less than 0.2 mm is adequate to neglect the effect of event migrations between the inside and outside of the regions. 

%%%%%%%%%%%%%%%%%%%%%%%%%%%%%%%%%%%%%%%%%%%%%%%%%%%%%%%%%%%%%%%%%%%
\subsection{Corrections}
\label{sec:Corrections}

\begin{figure*}[t]
  \centering
  \includegraphics[width=\textwidth]{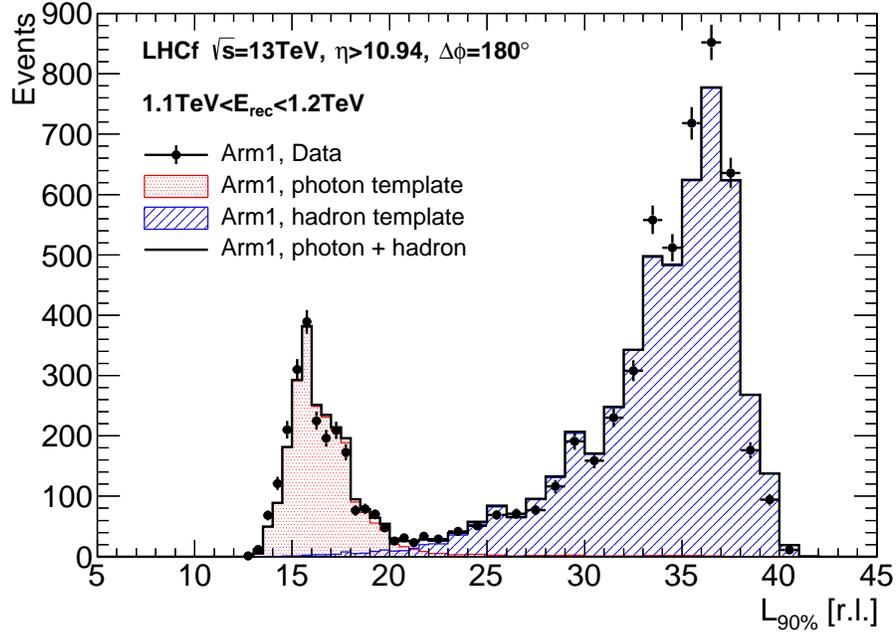}
  \caption{ 
  $L_{90\%}$ distribution in Arm1 for the events with the reconstructed energy between 1.1 and 1.2 TeV. 
  The black points represent the experimental data with statistical error bars. The red and blue coloured lines correspond to 
  the template distributions obtained from the MC simulation for photons and hadrons, respectively. 
  The black line represents the total of the template distributions. 
  These distributions were normalised by the results of the template fitting.   
  }
  \label{fig:L90}
 \end{figure*}

\begin{itemize}
\item Beam-related background 

The contribution of background events is due to interactions between the circulating beams and residual gas in the beam pipe. 
The background was estimated using the events associated with non-crossing bunches at IP1.
These events were generated purely from the beam-gas interactions, while the events associated with the colliding bunches were related to both the signal and background. The estimated background-to-signal ratio was less than 1\%; this ratio was subtracted from the measured cross-section.  
The difference in bunch intensity between colliding and non-colliding bunches was considered in the calculation.
Because of the limited statistics of the non-colliding bunch data, the correction was applied as an energy-independent factor;
nonetheless, the shape of the background spectrum is consistent with the shape of the signal.

\item PID correction

Corrections related to the PID selection, the inefficiency of photon selection and the contamination of hadrons, were performed using the template-fit method of the distribution of the PID estimator, $L_{90\%}$, defined as the longitudinal depth, in units of radiation length ($X_{0}$), at which the integral of the energy deposition in a calorimeter reached 90\% of the total.  
As a criterion of the selection of the photon component, we set an energy-dependent criterion $L_{90\%,thr}$, which defines the $L_{90\%}$ value to maintain a 90\% efficiency of photon selection in the MC simulations.  
Figure~\ref{fig:L90} presents the $L_{90\%}$ distribution of Arm1-Region A for the reconstructed energy range between 1.1 and 1.2 TeV. The red and blue lines in Fig.~\ref{fig:L90}, obtained from the MC simulation dataset of QGSJET~II-04, indicate the template distributions for the pure photon and pure hadron samples, respectively. These distributions were produced with normalization obtained from the template-fit result.
According to the template-fit results, the hadron contamination, typically 10\%, can be estimated as a function of energy and it is corrected together with the 90\% efficiency in the analysis.

\item Multi-hit correction

Because the mis-reconstruction of multi-hit events as single-hit events makes the measured spectra more complex, multi-hit events were rejected from the analysis. 
In order to identify multi-hit events, a lateral shower profile measured by the position-sensitive layers was fitted by an empirical function.
The difference in the goodness-of-fit between the single and double peak assumptions, the distance between two peaks, and the ratio between two peak heights were used to identify multi-hit events. 
These criteria were adjusted to achieve a high efficiency of multi-hit detection while maintaining a reasonably low incidence of single-hit-event mis-reconstructions as multi-hit events.

The consistency of the multi-hit identification efficiencies exhibited by the data and MC simulation was tested using `artificial' multi-hit event sets.
These artificial multi-hit events were created by merging two independent single-hit events. The combinations of single-hit events were selected to represent the distributions of photon-pair energies and hit-position distances in the true multi-hit events of QGSJET~II-04.
The same procedure was performed for the MC simulation also.
The multi-hit detection efficiency exceeds 85\% across the full energy range and reaches nearly 100\% above 2 TeV, while inconsistencies between the data and MC are less than approximately 5\% and 10\% for Arm1 and Arm2, respectively.
In the high-energy range, most of the multi-hit events are caused by photon pairs from $\pi^{0}$ decay.
In these events, the separation between photons is kinematically limited above 5.8\,mm.
This makes the identification of multi-hits simpler.

About 4\% of the total triggered events were identified as multi-hit events. 
Two corrections were applied to the measured cross-section:
\begin{enumerate}
\item `Multi-hit performance' correction:
  
  The contamination of multi-hit events misidentified as single-hits and the loss of single-hit events misidentified as multi-hits are corrected with an energy-dependent factor based on the MC dataset of QGSJET II-04. This correction factor depends mostly on the detector performance, while it depends weakly on the model chosen to generate the dataset.
  
\item `Multi-hit cut' correction:

  As the single-photon cross-section is measured by the detector, another correction factor based on the same MC dataset was applied to correct for the multi-hit cut and recover the inclusive production cross-section. This correction factor ranged within $\pm 50\%$, which was the largest contribution among the corrections and was strongly dependent on the choice of event-generation model in the MC simulation.
This is because the multi-hit rate is related to the cross-section of high-energy $\pi^{0}$ production, as discussed above.
\end{enumerate}
Both multi-hit corrections were performed inside the unfolding algorithm, which is described below.

\item Unfolding:

  We corrected for detector biases (as energy resolution and multi-hit effects) in the obtained cross-section by performing an unfolding technique based on the iterative Bayesian method~\cite{ref:unfolding} provided by the RooUnfold package~\cite{ref:RooUnfold}. The MC simulation dataset with $10^{8}$ inelastic collisions generated by the QGSJET~II-04 model was used as a training sample.  

\item Decay correction:
  
  The photons detected by the LHCf experiment mainly come from the decay of short-lifetime particles such as $\pi^0$ and $\eta$ mesons, which decay near the interaction point.
  Particles with a longer lifetime (such as $K^0$, $K^{\pm}$ and $\Lambda$) can decay along the beam pipe between the interaction point and detector and can contribute to the photon yield.
  In order to remove the contribution of long-lifetime particles, an energy-dependent correction was estimated with MC simulations by comparing the photon production cross-section at the interaction point with that after transportation along the beam pipe to the detector (i.e. after step `2' described in Sec.~\ref{sec:mc}).
  The correction reaches a maximum of about 15\% in the lowest-energy bin and becomes less than 1\% above 2 TeV.
  
\end{itemize}

%%%%%%%%%%%%%%%%%%%%%%%%%%%%%%%%%%%%%%%%%%%%%%%%%%%%%%%%%%%%%%%%%%%
\section{Systematic Uncertainties}

\begin{figure*}[t]
   \centering
   \includegraphics[width=\textwidth]{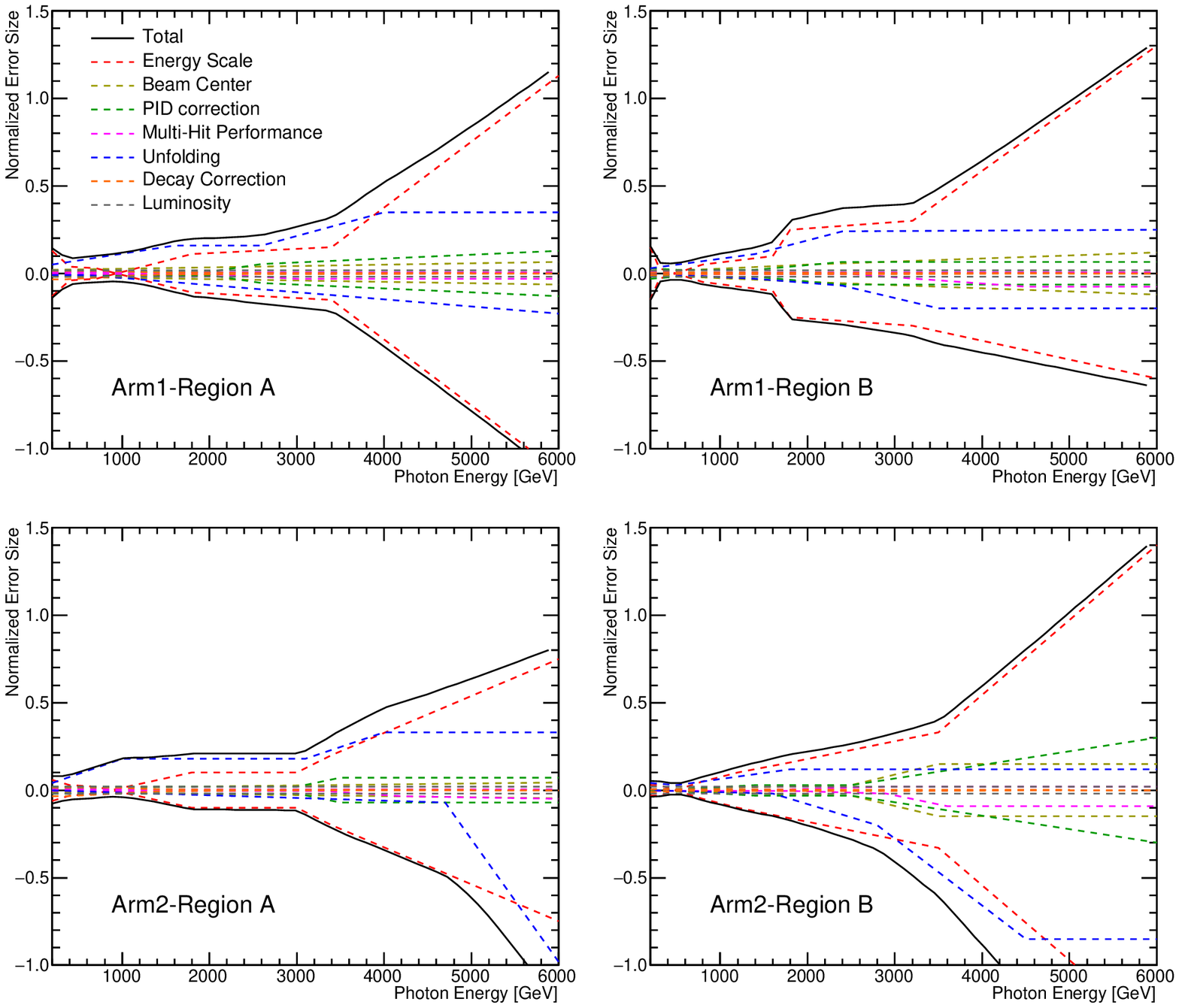}
  \caption{
  Systematic uncertainties of the photon production cross-section in the Arm1 (top) and Arm2 (bottom) analyses. 
  The left and right panels correspond to the results of the two analysis regions. 
  The coloured and dashed lines indicate the estimated systematic uncertainties after normalization with the mean values of the experimental data.
  The black line indicates the total systematic uncertainties calculated as quadratic summations of all the uncertainties. 
  }
  \label{fig:syserr}
 \end{figure*}

We considered the following contributions as systematic uncertainties of the measured production cross-section.  
Figure~\ref{fig:syserr} shows the estimated systematic uncertainties for each detector and each region as a function of photon energy.

\subsection{Energy scale}
\label{sec:sys_scale}
Energy scale errors are attributable to a) the absolute gain calibration of each sampling layer, b) uniformity, c) relative gain calibration of the photomultiplier tubes (PMTs) used for the readout of scintillator lights, and d) the Landau--Pomeranchuk--Migdal (LPM) effect~\cite{ref:LPM1,ref:LPM2}. The first two contributions were studied in beam tests and are described in Ref.~\cite{ref:BeamTest}. The third source of errors is related to the differences in the high-voltage configurations of PMTs between the beam tests and operation. The error was about 1.9\%.
The contribution to the error from the LPM effect was estimated as 0.7\% by comparing the detector responses upon activation and inactivation of the LPM effect in the detector simulation. 
The total energy-scale error, estimated from the quadratic summation of all contributions, was $\pm3.4\%$ for Arm1 and $\pm2.7\%$ for Arm2. The systematic uncertainty of the cross-section was estimated by shifting the energy scale within the errors.

\subsection{Beam-center stability}
The beam center, an important parameter for defining analysis regions, was calculated from the measured hit-map distribution of the hadronic shower events, which were selected such that $L_{90\%} > L_{90\%,thr}$. The fluctuations between subsequent data subsets were found to be of the order of 0.3 mm, which is greater than the statistical uncertainty of the mean beam-center measurements that used all the data in the Fill.
The systematic uncertainty associated with the beam-center determination was estimated by artificially moving the beam-center position by $\pm0.3~\mathrm{mm}$ on the x- and y-axes. 
The measured cross-section with the shifted beam-center positions was compared to the original cross-section and the variation was deemed to be the systematic uncertainty.

\subsection{PID}
\label{sec:sys_pid}
The contribution from the uncertainty on the fit of the $L_{90\%}$ distributions was negligible with respect to the statistical error of the cross-section.
The systematic uncertainty associated with the PID correction was estimated instead by changing the criterion for the choice of $L_{90\%,thr}$ to discriminate between photons and hadrons, as discussed above.
Instead of choosing $L_{90\%,thr}$ to obtain a 90\% photon selection efficiency, PID selection and correction were also performed using the threshold values that produced photon-selection efficiencies of 85\% and 95\%.
The 85\%--95\% limits were chosen in order to maintain the `efficiency $\times$ purity' product above 75\% in the full energy range. 
We compared the measured cross-section after correction and determined the systematic uncertainty from the relative deviation from the original cross-section.

\subsection{Multi-hit identification efficiency}
\label{sec:sys_mul}

The correction factors attributable to the `multi-hit performance' were obtained from the MC simulation.
Thus, we tested the consistency of the multi-hit identification efficiencies exhibited by the data and the MC simulation by using the artificial multi-hit event sets, as previously described in Sec.~\ref{sec:Corrections}.
The systematic uncertainty on the production cross-section was calculated by multiplying the relative error of the multi-hit identification efficiency (i.e. the discrepancy between the data and MC simulation) by the ratio of multi-hit events to single-hit events.

\subsection{Unfolding}
\label{sec:sys_unf}

It was discovered that the interaction model dependency of the `multi-hit cut' correction factors, computed from the training sample, was 
the main source of systematic uncertainty in the cross-section unfolding process.
EPOS-LHC predicted a higher multiplicity of photons than QGSJET~II-04. Thus, a larger correction factor was expected in EPOS-LHC than in QGSJET~II-04. We performed cross-section unfolding with a training sample of $5\times10^{7}$ inelastic collisions generated by EPOS-LHC. The relative difference between the QGSJET~II-04 and EPOS-LHC results was chosen as the systematic uncertainty associated with the unfolding.

\subsection{Decay correction}
\label{sec:sys_decay}

The systematic uncertainty related to the correction for the decay of long-lifetime particles was estimated as the maximum relative fluctuation between the corrections predicted by the EPOS-LHC, QGSJET~II-04, DPMJET~3.06, SIBYLL~2.3, and PYTHIA~8.212 models.

%%%%%%%%%%%%%%%%%%%%%%%%%%%%%%%%%%%%%
\section{Results}
\label{sec:Results}
\begin{figure*}[t]
   \centering
   \includegraphics[width=\textwidth]{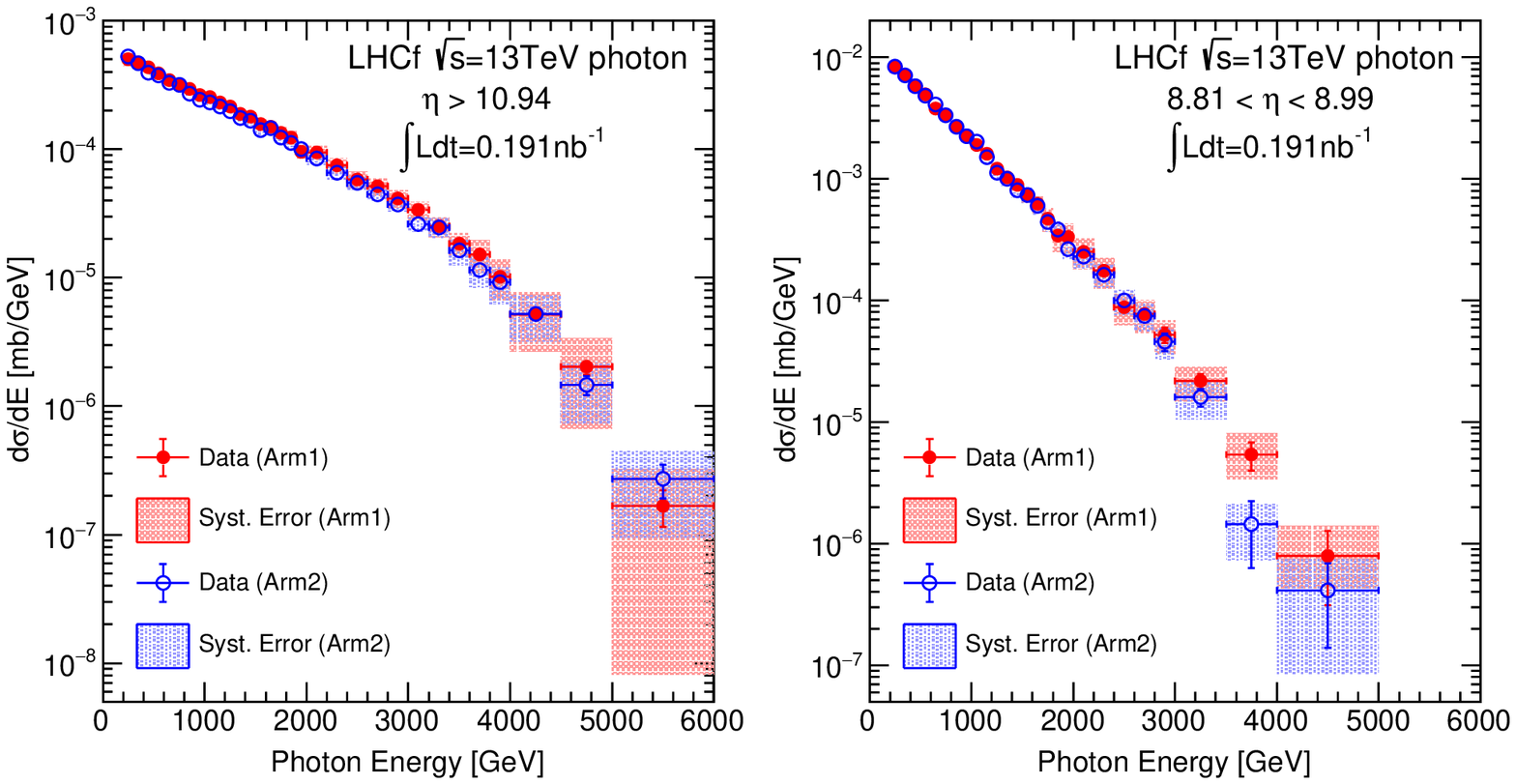}
  \caption{
  Photon production cross-section measured by the Arm1 (red filled circle) and Arm2 (blue open circle) detectors. 
  The left figure presents the results for $\eta\,>\,10.94$, which covers the zero-degree collisions angle. 
  The right figure presents those for $8.81\,<\,\eta\,<\,8.99$, which corresponds to the fiducial area 
  in the large calorimeters of the detectors. 
  The bars and hatched areas correspond to the statistical and systematic uncertainties, respectively. 
  Only uncorrelated systematic uncertainties between Arm1 and Arm2 are considered in these plots. 
  }
  \label{fig:arm12}
 \end{figure*}

Figure~\ref{fig:arm12} presents the photon production cross-section measured by the Arm1 and Arm2 detectors.
The error bars and hatched areas indicate the statistical and systematic uncertainties, respectively. 
In this comparison of the results of the two detectors, the detector-correlated systematic uncertainties due to the luminosity, unfolding, and decay correction were not considered.
We found a general agreement, within the given uncertainties, between the results of the two detectors.

\begin{figure*}[t]
   \centering
   \includegraphics[width=\textwidth]{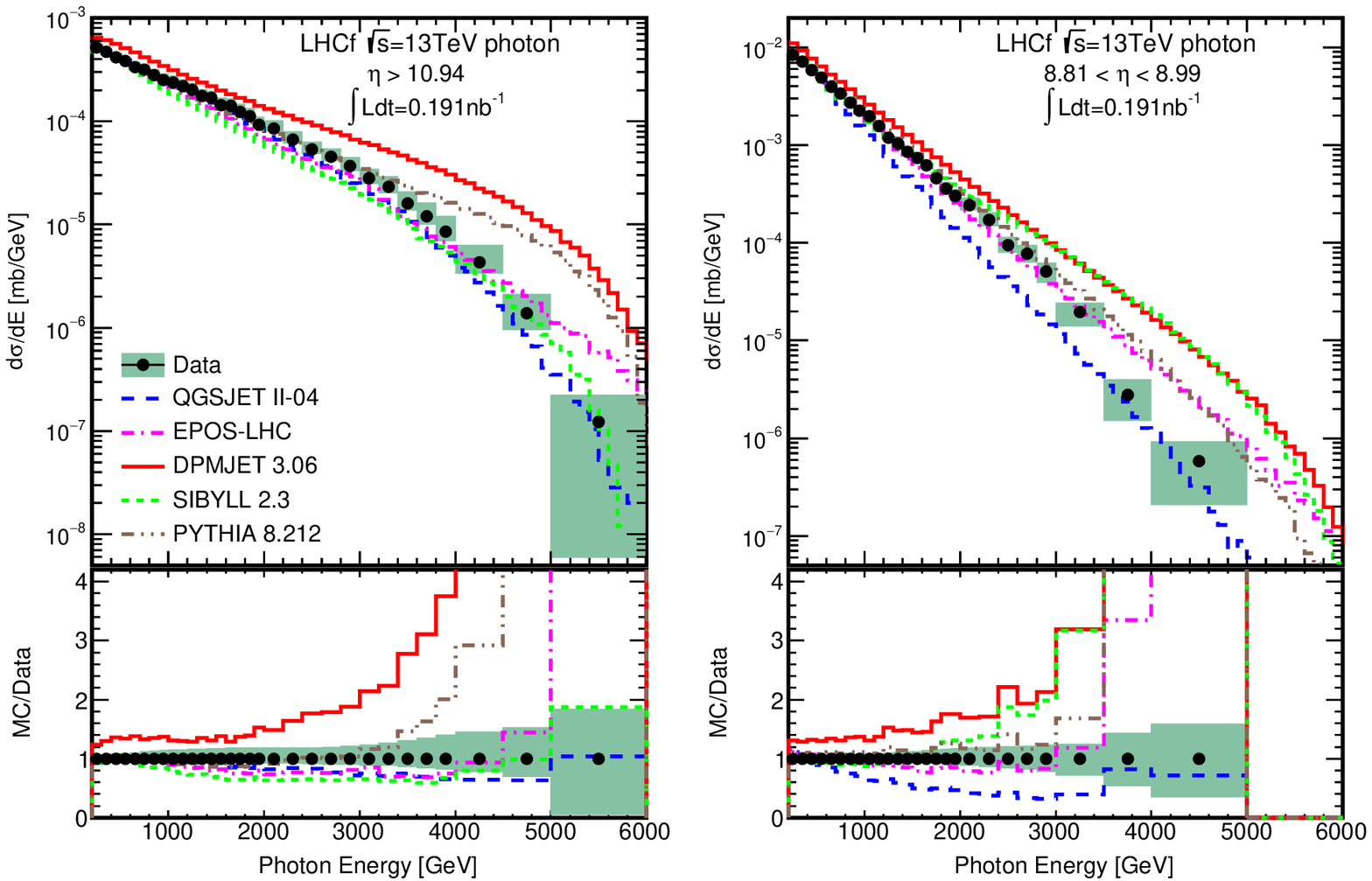} 
  \caption{ 
  Comparison of the photon production cross-section obtained from the experimental data and MC predictions. 
  The top panels show the cross-section and the bottom panels show the ratio of MC predictions to the data. 
  The shaded areas indicate the total uncertainties of experimental data including the statistical and systematic uncertainties. 
  }
  \label{fig:spectra}
 \end{figure*}

We combined the results using the same method as the analysis presented in Ref.~\cite{ref:7TeV-Pi0}. This approach assumed that the systematic uncertainties of the energy scale, PID correction, performance of multi-hit identification, and beam position exhibited both bin-by-bin correlation and Arm1-Arm2 non-correlation.
The other systematic uncertainties---luminosity, unfolding, and decay correction---were assumed to be fully correlated between Arm1 and Arm2. These uncertainties were added quadratically to the combined results.  
The upper panels of Fig.~\ref{fig:spectra} show the combined cross-section with the predictions of the hadronic interaction models, QGSJET~II-04, EPOS-LHC, DPMJET~3.06, SIBYLL~2.3, and PYTHIA~8.212. The shaded areas indicate the total statistical and systematic uncertainties, which were calculated using the combining method. The bottom panels show the ratio of MC predictions to the experimental results.
In the pseudorapidity region $\eta\,>\,10.94$, the QGSJET~II-04 and EPOS-LHC models show the best agreement overall with the data. PYTHIA~8.212 shows good agreement with the data from the lowest-energy bin to near the 3 TeV bin, although it clearly predicts a higher cross-section than the data in the energy region greater than 3 TeV. 
DPMJET~3.06 and SIBYLL~2.3 predict fluxes higher and lower, respectively, than the data in most of the energy range. 
In the pseudorapidity region $8.81\,<\,\eta\,<\,8.99$, results from the EPOS-LHC and PYTHIA~8.212 models show good agreements with the data except at the high-energy end above 3 TeV. 
QGSJET~II-04 and DPMJET~3.06 predict fluxes lower and higher, respectively, than the data. 
SIBYLL~2.3 exhibits a different trend from the result in $\eta >$ 10.94, predicting a higher cross-section than the data in the energy range above $>$ 1.5 TeV. 
This result is related to the fact that SIBYLL~2.3 predicts a larger mean value of $p_{\mathrm{T}}$ for photons than both the data and other models. 
 
The general trends demonstrated by the data and MC simulations resemble the results obtained from proton--proton collisions at \sqrts \,=\, 7\,TeV in Ref.~\cite{ref:7TeV-photon}, which showed the measured energy spectra for forward photons in the same pseudorapidity regions compared to MC predictions from QGSJET~II-03, EPOS~1.99, SIBYLL~2.1, DPMJET~3.04, and PYTHIA~8.145. Except for DPMJET~3.04, these models are older versions than those to which Fig.~\ref{fig:spectra} refers. The updates to these models and differences in collision energy do not produce significant changes in the forward photon production cross-section in the QGSJET~II and EPOS models. Thus, the detailed differences in the results from \sqrts \,=\, 7~TeV and \sqrts \,=\, 13~TeV may correspond to the differences between the $p_{\mathrm{T}}$ coverages.

%%%%%%%%%%%%%%%%%%%%%%%%%%%%%%%%%%%%%
\section{Summary}
The LHCf experiment measured the production cross-section of forward photons at $\eta\,>\,10.94$ and $8.99\,>\,\eta\,>\,8.81$ with proton--proton collisions at \sqrts =\,13\,TeV. 
The two LHCf detectors, Arm1 and Arm2, produced consistent results, which were combined while considering their statistical and systematic uncertainties. 
The final results were compared to the MC predictions obtained from several hadronic interaction models: QGSJET~II-04, EPOS-LHC, DPMJET~3.06. SIBYLL~2.3, and PYTHIA~8.212. 
Among these models, EPOS-LHC showed the best agreement with the experimental data. 
QGSJET~II-04 showed good agreement with the data for $\eta\,>\,10.94$ but predicted a lower flux than the data for $8.99\,>\,\eta\,>\,8.81$.  
PYTHIA~8.212 showed a higher cross-section than the data in the energy region above 3 TeV. 

No MC models matched the experimental data perfectly. The differences between the data and MC models were attributable to a less-than-complete understanding of the soft hadronic interaction processes. Common operations of the LHCf with the ATLAS experiment, in which the detector covers the central region of IP1, were performed in 2015. Detailed studies with event-by-event information measured by ATLAS will help us better understand the production of photons in the forward region~\cite{ref:Diffractive_Zhou}.

%%%%%%%%%%%%%%%%%%%%%%%%%%%%%%%%%%%%%
\section*{Acknowledgments}
We thank the CERN staff and ATLAS Collaboration for their essential contributions to the successful operation of LHCf.  
This work was partly supported by JSPS KAKENHI(Grant Numbers JP26247037, JP23340076) and the joint research program of the Institute for Cosmic Ray Research (ICRR), University of Tokyo. This work was also supported by Istituto Nazionale di Fisica Nucleare (INFN) in Italy.
Parts of this work were performed using the computer resource provided by ICRR (University of Tokyo), CERN and CNAF (INFN).

%%%%%%%%%%%%%%%%%%%%%%%%%%%%%%%%%%%%%
\clearpage

\appendix
\section{Cross-section table}

\begin{table}[!htbp]
  %\scriptsize
  \footnotesize
  \begin{center}
    \begin{tabular}{|c|c|c|c|}
      \hline
      \multicolumn{2}{|c|}{\boldmath$\eta > 10.94$} & \multicolumn{2}{|c|}{\boldmath$8.81 < \eta < 8.99$} \\
      \hline
      \textbf{Energy [GeV]} & \boldmath$d\sigma/dE$ \textbf{[mb/GeV]} & \textbf{Energy [GeV]} & \boldmath$d\sigma/dE$ \textbf{[mb/GeV]} \\
      \hline
      200--300	& $(5.21^{+0.32}_{-0.25}) \times 10^{-4}$ &  200--300	& $(8.42^{+0.35}_{-0.19}) \times 10^{-3}$ \\
      300--400	& $(4.72^{+0.33}_{-0.21}) \times 10^{-4}$ &  300--400	& $(7.16^{+0.29}_{-0.12}) \times 10^{-3}$ \\
      400--500	& $(4.15^{+0.34}_{-0.17}) \times 10^{-4}$ &  400--500	& $(5.84^{+0.25}_{-0.09}) \times 10^{-3}$ \\
      500--600	& $(3.82^{+0.36}_{-0.15}) \times 10^{-4}$ &  500--600	& $(4.93^{+0.23}_{-0.09}) \times 10^{-3}$ \\
      600--700	& $(3.36^{+0.36}_{-0.13}) \times 10^{-4}$ &  600--700	& $(3.97^{+0.21}_{-0.09}) \times 10^{-3}$ \\
      700--800	& $(3.16^{+0.38}_{-0.11}) \times 10^{-4}$ &  700--800	& $(3.38^{+0.21}_{-0.09}) \times 10^{-3}$ \\
      800--900	& $(2.82^{+0.37}_{-0.10}) \times 10^{-4}$ &  800--900	& $(2.73^{+0.19}_{-0.08}) \times 10^{-3}$ \\
      900--1000	& $(2.51^{+0.36}_{-0.09}) \times 10^{-4}$ &  900--1000	& $(2.27^{+0.17}_{-0.08}) \times 10^{-3}$ \\
      1000--1100	& $(2.39^{+0.36}_{-0.09}) \times 10^{-4}$ & 1000--1100	& $(1.98^{+0.17}_{-0.08}) \times 10^{-3}$ \\
      1100--1200	& $(2.19^{+0.34}_{-0.09}) \times 10^{-4}$ & 1100--1200	& $(1.57^{+0.15}_{-0.07}) \times 10^{-3}$ \\
      1200--1300	& $(2.01^{+0.33}_{-0.09}) \times 10^{-4}$ & 1200--1300	& $(1.19^{+0.12}_{-0.06}) \times 10^{-3}$ \\
      1300--1400	& $(1.76^{+0.30}_{-0.08}) \times 10^{-4}$ & 1300--1400	& $(1.03^{+0.11}_{-0.05}) \times 10^{-3}$ \\
      1400--1500	& $(1.68^{+0.29}_{-0.08}) \times 10^{-4}$ & 1400--1500	& $(8.58^{+1.01}_{-0.49}) \times 10^{-4}$ \\
      1500--1600	& $(1.44^{+0.26}_{-0.08}) \times 10^{-4}$ & 1500--1600	& $(7.43^{+0.95}_{-0.45}) \times 10^{-4}$ \\
      1600--1700	& $(1.42^{+0.26}_{-0.08}) \times 10^{-4}$ & 1600--1700	& $(6.18^{+0.86}_{-0.43}) \times 10^{-4}$ \\
      1700--1800	& $(1.24^{+0.23}_{-0.08}) \times 10^{-4}$ & 1700--1800	& $(4.61^{+0.72}_{-0.39}) \times 10^{-4}$ \\
      1800--1900	& $(1.13^{+0.21}_{-0.08}) \times 10^{-4}$ & 1800--1900	& $(3.60^{+0.60}_{-0.36}) \times 10^{-4}$ \\
      1900--2000	& $(9.28^{+1.75}_{-0.66}) \times 10^{-5}$ & 1900--2000	& $(3.02^{+0.54}_{-0.32}) \times 10^{-4}$ \\
      2000--2200	& $(8.56^{+1.59}_{-0.62}) \times 10^{-5}$ & 2000--2200	& $(2.43^{+0.44}_{-0.28}) \times 10^{-4}$ \\
      2200--2400	& $(6.66^{+1.24}_{-0.53}) \times 10^{-5}$ & 2200--2400	& $(1.71^{+0.34}_{-0.23}) \times 10^{-4}$ \\
      2400--2600	& $(5.33^{+1.01}_{-0.46}) \times 10^{-5}$ & 2400--2600	& $(9.47^{+1.99}_{-1.54}) \times 10^{-5}$ \\
      2600--2800	& $(4.55^{+0.90}_{-0.43}) \times 10^{-5}$ & 2600--2800	& $(7.74^{+1.69}_{-1.43}) \times 10^{-5}$ \\
      2800--3000	& $(3.70^{+0.79}_{-0.37}) \times 10^{-5}$ & 2800--3000	& $(5.06^{+1.20}_{-1.14}) \times 10^{-5}$ \\
      3000--3200	& $(2.81^{+0.66}_{-0.32}) \times 10^{-5}$ & 3000--3500	& $(1.96^{+0.48}_{-0.56}) \times 10^{-5}$ \\
      3200--3400	& $(2.31^{+0.62}_{-0.28}) \times 10^{-5}$ & 3500--4000	& $(2.78^{+1.20}_{-1.27}) \times 10^{-6}$ \\
      3400--3600	& $(1.60^{+0.49}_{-0.22}) \times 10^{-5}$ & 4000--5000	& $(0.58^{+0.34}_{-0.38}) \times 10^{-6}$ \\
      3600--3800	& $(1.21^{+0.43}_{-0.20}) \times 10^{-5}$ & & \\
      3800--4000	& $(8.52^{+3.44}_{-1.55}) \times 10^{-6}$ & & \\
      4000--4500	& $(4.31^{+1.97}_{-0.93}) \times 10^{-6}$ & & \\
      4500--5000	& $(1.39^{+0.73}_{-0.43}) \times 10^{-6}$ & & \\
      5000--6000	& $(1.22^{+1.02}_{-1.16}) \times 10^{-7}$ & & \\
      \hline
    \end{tabular}
    \caption{Differential photon production cross-section $d\sigma/dE$ [mb/GeV] for each energy bin and pseudorapidity range.
    Upper and lower total uncertainties are also reported.}
    \label{tab:results}
  \end{center}
\end{table}

\clearpage

%%%%%%%%%%%%%%%%%%%%%%%%%%%%%%%%%%%%%
\section*{References}

%\bibliography{mybibfile}

\end{document}